\documentclass[]{pasa}
\usepackage{graphicx}
\usepackage{epstopdf}
 
\usepackage{aas_macros}
\usepackage{hyperref} 
\hypersetup{colorlinks,citecolor=blue,linkcolor=blue,urlcolor=blue}
\usepackage[authoryear]{natbib}

\title[Pop II]{Where is Population II?}
\author{J.~Mould$^{1,2}$, F.~Bianchini$^3$, Duncan ~A.~Forbes$^1$, C.~L.~Reichardt$^3$} 
\begin{document}
%
\maketitle
\leftline{$^1$CAS, Swinburne University, Vic 3122, Australia}
\leftline{$^2$ARC CoE for All-Sky Astrophysics, CAASTRO}
\leftline{$^3$School of Physics, Melbourne University, Vic 3010}
\begin{abstract}
The use of roman numerals for stellar populations represents a classification
approach to galaxy formation which is now well behind us. Nevertheless, the
concept of a pristine generation of stars, followed by a protogalactic era, and finally the mainstream stellar population is a plausible starting point for testing our physical understanding of early star formation. 
This will be observationally driven as never before in the coming decade. 
In this paper, we search out observational tests of an idealized coeval and homogeneous distribution of population II stars.
We examine the spatial distribution of quasars, globular clusters, and the integrated free electron density of the intergalactic medium, in order to test the assumption of homogeneity. Any $real$ inhomogeneity implies a population II that is not coeval.
\end{abstract}




\section{Introduction}
`{\it Where is Population III?}' is an important question \citep{Ishiyama2016} and often asked rhetorically. 
The question asked in this paper -- whether population II stars are coeval and homogeneously distributed --  is no less important for different reasons.
Whether population II is present in all galaxies and is coeval everywhere places constraints on cosmology and galaxy formation and could impact theories such as timescape cosmology \citep{Wiltshire2009} or the multiverse \citep{rees2001,linde}. 
More generally, comparing the observed age of Population II stars to the age of the Universe constrains the cosmological constant and is one of the three pillars supporting the standard model of cosmology, the others being the anisotropy of the cosmic microwave background \citep[CMB]{planck_params} and the Hubble diagram of type Ia supernovae \citep{Turner2001}. 

In this work, we examine the spatial distribution of tracers of population II stars: Globular Clusters (GCs) at low-redshift (\S\ref{sec:gc}) and at high-redshift, quasi-stellar objects (QSOs; \S\ref{sec:qso}) ) and the optical depth to reionization along different lines of sight (\S\ref{sec:tau}). 
Motivated by the apparent isotropy in GCs and anisotropy in the high-z QSO distribution, we conclude by discussing the possibility of a non-coeval population II objects. 


\section{Globular Clusters -- local tracers of old Pop II objects}\label{sec:gc}

Globular clusters (GCs) have long been recognised as one of the oldest known stellar systems. Although many are probably destroyed over cosmic time, some survive till today, being ubiquitous around large galaxies and occasionally hosted by dwarf galaxies (see \cite{Brodie2006} for a review). Globular cluster systems usually reveal a bimodal colour distribution, which is largely due to a $\sim$ 1 dex difference in their metallicity \citep{Usher2012}. These blue/metal-poor and red/metal-rich subpopulations of GCs may also have a small difference in their mean age with the red/metal-rich GCs being younger $\sim$1 Gyr  (although formally it is difficult to rule out coeval ages). Current absolute age estimates for Milky Way GCs, based on main sequence and white dwarf cooling fitting, indicates that the blue/metal-poor GCs were formed around 12.5 Gyr ago, however the uncertainties on this value place the metal-poor GCs well within the epoch of reionisation (i.e. $z >$ 6, age $>$ 12.8 Gyr).  Extragalactic GCs are less well constrained but are consistent with the old ages of their Galactic counterparts \citep{Strader2005,Wagner-Kaiser2017}.
For a summary of GC ages from observational constraints, and predictions from various simulations; see \cite{Forbes2015} and \cite{Mould1998}. More recently, a simulation of GC systems around Milky Way like galaxies by \cite{pfeffer2017} predicts GCs to start forming before a redshift of 6 and continue to earlier times. HST observations,  exploiting the amplification of strong lensing, have identified possible proto-GCs at $z = 6.1$ \citep{Vanzella2017}. 
Another important feature of metal-poor GCs is that they appear to have a `metallicity floor', i.e. no GC has been observed with a metallicity [Z/H] below -2.5 (e.g. \citet{Usher2012}). The reason for this minimum metallicity is not yet fully understood but may be a combination of pre-enrichment by Pop III stars \citep{Beasley2003} and gas cooling by metals and molecules \citep{Glover2014}.

\begin{figure*}
\includegraphics[scale=0.7, angle=-90]{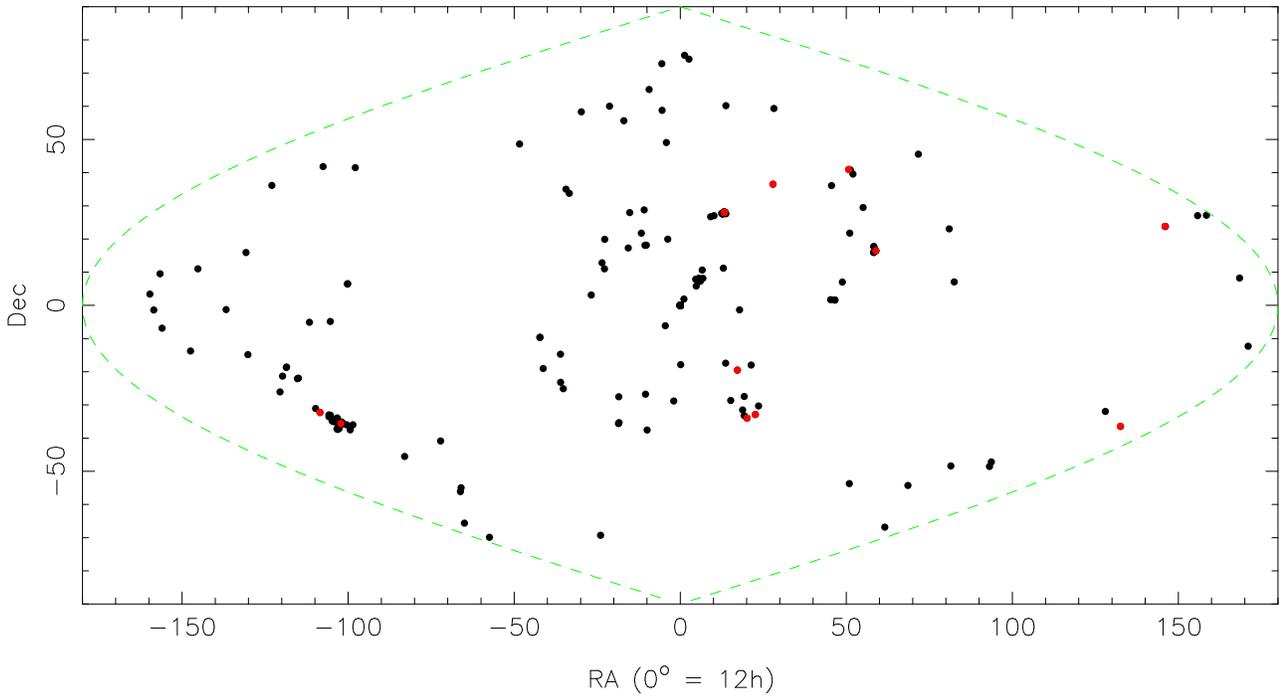}
\vspace*{-2 truecm}  
\caption{\label{fig:gc}
The shell of globular cluster systems beyond 20 Mpc. Globular clusters
beyond 100 Mpc are shown in red, with the most distant being globular clusters located in Abell 1689 at 790 Mpc or 2.5 Glyr.
 Data are from the compilation of \citet{Harris2013}.}
\end{figure*}

In Figure \ref{fig:gc}, we show the space distribution of known GC systems from the compilation of \cite{Harris2013}. Although this compilation does not separate the two subpopulations, previous work has shown the blue/metal-poor subpopulation is found in all GC systems unlike the red/metal-rich subpopulation which is increasingly absent in lower mass galaxies \cite{Peng2006}. Thus the distribution can be taken as a proxy for the old, population II metal-poor GCs. 
The GCs are isotropically distributed on the sky modulo the regions associated with the Galactic disk. 

\section{Quasars  -- tracing early Pop II objects} \label{sec:qso}

\begin{figure*}
\includegraphics[scale=0.7, angle=-90]{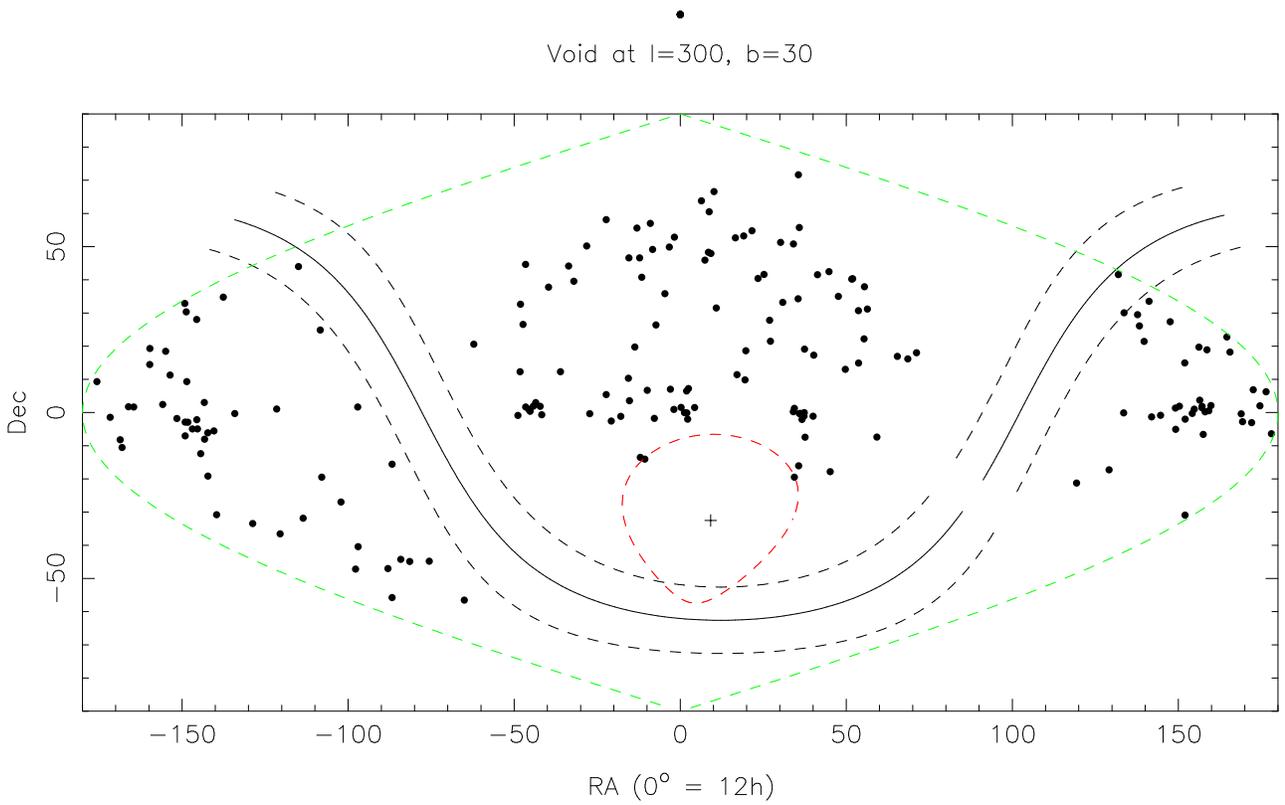}  
\caption{\label{fig:qso}
QSOs with $z >$ 6 distributed on the sky. RA horizontal, Dec. vertical. We show the galactic plane $\pm$ 10$^\circ$ and in green the boundaries of the projection. A 30$^\circ$ radius void is shown in red.}
\end{figure*}

Given a conservative lower limit of 12 Gyrs for the age of Population II, these stars formed at $z >6$  according to the standard model of cosmology with $Planck$ collaboration parameters \citep{planck_params}. 
 Unfortunately it is as yet infeasible to observe directly low-mass objects in this epoch, so we must rely on more luminous tracers of the star formation activity. 
We turn to quasars because  (i) they are observationally feasible, and (ii) there is a well-discussed link between quasar activity and star formation rates in high-redshift galaxies. 
For instance, a positive correlation between quasar luminosity and far-IR luminosity has been seen previously \citep{Wang2011,Omont2013,Venemans2016}.
The spatial distribution of quasars at $z>6$ in the Million Quasar Catalog\footnote{ https://heasarc.gsfc.nasa.gov/w3browse/all/milliquas.html} \citep{Flesch2015}
 is shown in Figure \ref{fig:qso}. 
 Two features catch the eye: the Galactic plane, and some apparent voids. 
The dearth of high-$z$ QSOs in the Galactic Plane is easily explained as obscuration by the Milky Way. 
The largest void is at 12 hours right ascension and -60$^\circ$ declination. 
 It appears to be of significant size ($\sim$1 Gpc in diameter), but this is an upper limit imposed by selection effects. 
 The reasons for this apparent southern void are not understood, but likely fall into one of two camps. 
 First, the void might be due to an unaccounted for selection effect related to the union of the heterogenous surveys that go into the Million Quasar Catalog. 
 Second and more interestingly, the void could be real. 
 If so, the void could point to large-scale anisotropies in the SFR at high-redshift in conflict with the predictions of the standard cosmological model.

\section{A hypothetical void}\label{sec:tau}

For the sake of argument let us suppose that the void in Figure 1 were real. This is a useful hypothesis to entertain as, if and when observers fill it, constraints on large scale inhomogeneities $(\S 1)$ will find application to cosmology. 
A Gpc-scale void in $z >$ 6 QSOs begs the question: what triggers the formation of supermassive black holes \citep{Wang2017}? 
If reflected in the star-formation rate, as suggested by the established connection between QSO activity and SFR at lower redshifts, this void also poses a new question: 
could the reionization of the Universe by the first stars be delayed on such enormous scales?

We can test this hypothesis even before the advent of 21 cm experiments  by studying the CMB anisotropies in the direction of the hypothesized void.  
Measurements of the CMB power spectra have already led to constraints on the average optical depth to reionization and thus the timing of reionization. 
There are two main effects. 
First, the optical depth screens the primordial CMB anisotropies and suppresses the power by a factor $e^{-2\tau}$ at all scales smaller than the horizon at the time of reionization. 
Unfortunately for measuring the mean optical depth,  this signature is strongly  degenerate with the amplitude of the primordial scalar perturbations power spectrum, $A_s$.  
Second,  Thomson scattering between the  CMB quadrupole and the free electrons from reionization generate linear polarization  at scales larger than the horizon size at the epoch of reionization. 
This signal shows up in the CMB $E$ and $B-$mode polarization angular power spectra as a ``bump'' at low multipoles, with the power scaling as $\tau^2$. 
The reionization bump is not degenerate with other cosmological parameters, but 
careful cleaning of the galactic foregrounds is required at  these angular scales, $\ell \lesssim 10$. 
The best constraints to date on the mean optical depth come from the \textit{Planck} satellite, $\tau = 0.058 \pm 0.012$, suggesting that reionization occurred around $z_{\rm re} \sim 8$ \citep{planck_reio}. Patchy reionization has
been explored by \citet{roy18}.

\subsection{Can we detect a lower optical depth towards the hypothetical void?}

We can use current observations to go beyond the mean optical depth across the full sky. 
We expect reionization to be a somewhat inhomogeneous process, and this inhomogeneity will cause variations in the optical depth integrated along different lines of sight. 
Broadly speaking, a larger value of $\tau$ in a given direction implies a higher $z_{\rm re}$ and thus an earlier onset of star and galaxy formation; $\tau$ = 0 implies no reionization at all.  
These variations in the optical depth will imprint a non-Gaussian signature in the CMB anisotropies. 
The non-Gaussianity can be detected with higher-order statistics of the map, as was done by \cite{Gluscevic2013} and \cite{Namikawa2017}. 
In the future, these methods will be able to create maps of the optical depth across the sky \citep{dvorkin2009}. 

In this work, we construct a simple estimator for the optical depth in the direction of the void and other similarly-sized patches on the sky. 
Under the reasonable assumption that the primordial amplitude $A_s$ is constant, the power spectrum in a given direction will relate to the full-sky CMB spectrum according to $\hat{C}^{TT}_{\ell}(\hat{\mathbf{n}}) = e^{-2(\tau(\hat{\mathbf{n}}) - \bar{\tau})}C_{\ell}^{TT,\,\rm full\,sky}$. 
Thus variations in the power spectrum amplitude across the sky reflect variations in the optical depth $\tau$. 
We use this idea to construct an estimator for the variation in the optical depth, $\delta\tau(\hat{\mathbf{n}}) = \tau(\hat{\mathbf{n}}) - \bar{\tau}$, as: 
\begin{equation}
\label{eq:tau_estimator}
\delta\tau(\hat{\mathbf{n}}) = -\frac{1}{2}\ln \left( \frac{\sum_{\ell}\frac{\ell(\ell+1)}{2\pi}C^{TT}_{\ell}(\hat{\mathbf{n}})}{\sum_{\ell}\frac{\ell(\ell+1)}{2\pi}C^{TT, \,\rm full\,sky}_{\ell}}\right). 
\end{equation}
Details of how we calculate the power spectra and apply this estimator can be found in the appendix. 
If the lack of high-redshift quasars in one direction is due to a delay in the onset of star formation and reionization, we would expect to see a lower optical depth and thus more CMB anisotropy power in that direction. 

\begin{figure}
\includegraphics[width=\columnwidth]{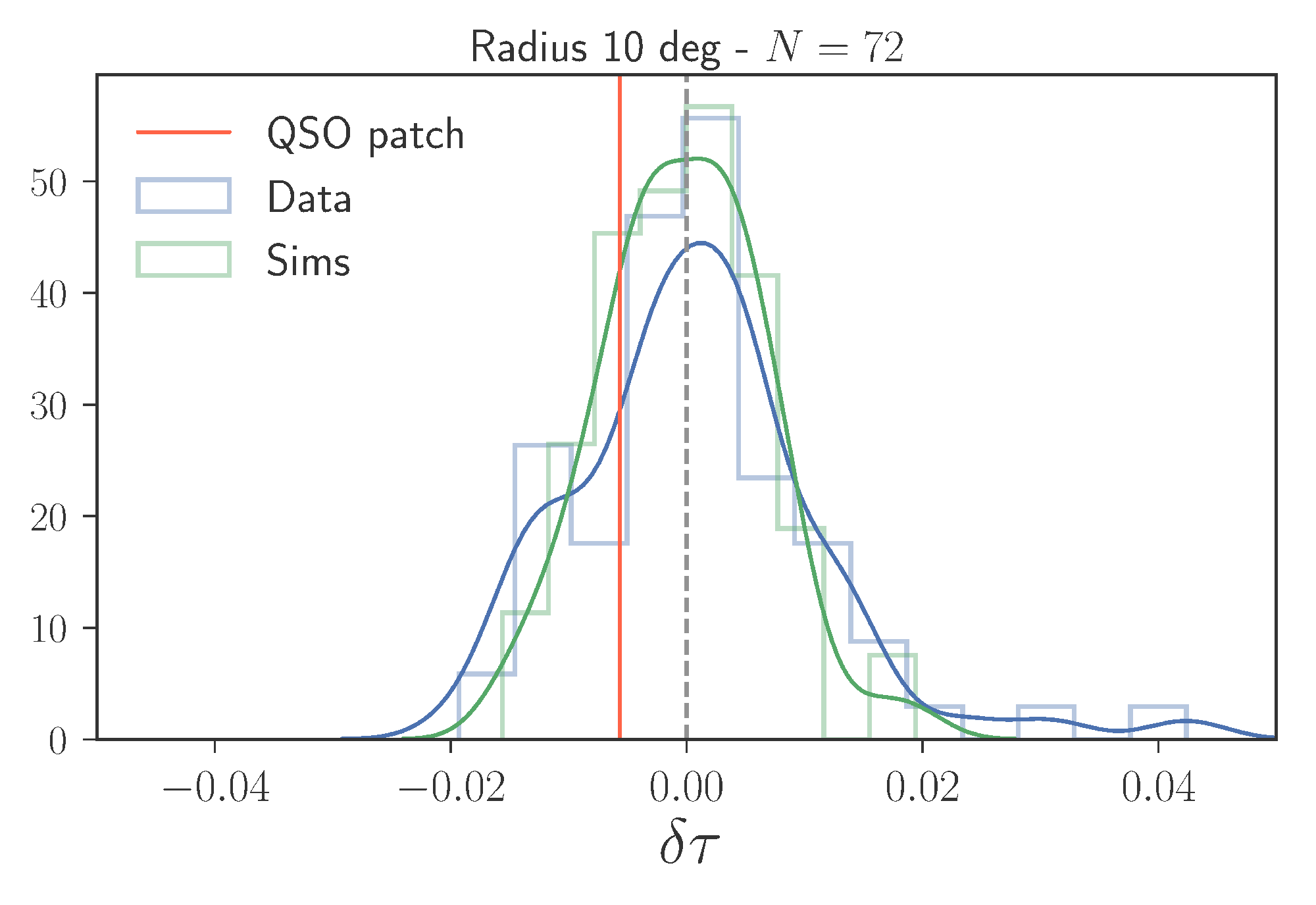}
\caption{Histograms of the recovered optical depth fluctuations (with respect to the sky averaged value) in 72 randomly distributed circular patches of 10 degree radius. Blue colour shows the results for the \textit{Planck} \texttt{SMICA} map while results based on the simulated map are shown in green. The vertical red line denotes the inferred value of $\tau$ in the hypothetical QSO void. The solid smooth lines represent kernel density estimates to the underlying distribution functions.}
\label{fig:tau_hist}
\end{figure}

The hypothesis that the QSO void is associated with a late reionization region therefore predicts  a negative $\delta\tau$ in the direction of the void. 
  To test this prediction, we estimate the optical depth over the $z \sim 6$ QSO void and also over a number of randomly positioned, same sized fields.
The results are shown in Fig.~\ref{fig:tau_hist}. 
  In the plot, we also compare the distribution of the inferred $\tau$ values at the same locations for  simulated noisy CMB temperature maps that do not include an inhomogeneous reionization signal. The first thing we notice is that the $\tau$ distribution found for real data appears to be slightly more non-Gaussian than the one found for simulations. 
  In particular, the distribution shows a positive $\tau$ tail, suggestive of early reionization in some areas. 
  We have tried varying the multipole range or reducing the mask radius to 5$^\circ$, and find the positive tail in all cases although its magnitude changes. 
Interestingly, the relative optical depth found in the hypothetical QSO void is negative as would be expected for a region with delayed reionization, however the magnitude is not statistically significant. We have also varied the radius of the extracted region without significant change.

Looking to the future, observations of 21 cm emission from neutral hydrogen should prove useful for tracking a more 
detailed reionization history and structure \citep{Jacobs2016,roy18}. 
Deep surveys continue to discover 
new galaxies at high redshifts (e.g. $z > 10$, \cite{Oesch2016}), and the James Webb Space Telescope will provide even greater
 detection sensitivity for distant galaxies near the beginning of reionization \citep{Wang2017}. WFIRST may see and record the history of the formation of GC systems \citep{Renzini2017}.
Improved constraints on $\tau$ are also expected from future high sensitivity observations of CMB polarization signals, such as CMB-S4 \citep{cmbs4} or the proposed CMBPol satellite \citep{Zaldarriaga2008}.  
\cite{Su2011} have considered inhomogeneous reionization and conclude that experiments such as CMBPol will be able to achieve a detection.

\section{Conclusion}
The apparent homogeneity in the extragalactic globular cluster distribution sets a lower bound on the radius of possible voids in population II of $\sim$80 Mpc in diameter (Appendix B). This is larger than the expectations of patchy reionization \citep{roy18,dragons}. The apparent void in the high $z$ QSO distribution provides an upper limit of order a Gpc.

There are a number of avenues by which inhomogeneities in Population II can be detected and confirmed. These include deeper all sky mapping of QSOs with $z >$ 6 (e.g. Taipan, \citet{taipan}), deep Hubble Space Telescope imaging of rich globular cluster systems at $z >$ 0.1, polarisation mapping of the CMB and imaging of reionization at 21 cm. This in turn will constrain some nonstandard inhomogeneous cosmologies, such as timescape and the multiverse.

\bibliographystyle{pasa-mnras}
\bibliography{trythis}

\begin{thebibliography}{}
\makeatletter
\relax
\def\mn@urlcharsother{\let\do\@makeother \do\$\do\&\do\#\do\^\do\_\do\%\do\~}
\definecolor{darkblue}{rgb}{0,0,0.597656}
\def\mndoi{\begingroup\mn@urlcharsother \@ifnextchar [ {\mndoi@} {\mndoi@[]}}
\def\mndoi@[#1]#2{\def\@tempa{#1}\ifx\@tempa\@empty \href
  {http://dx.doi.org/#2} {\textcolor{darkblue}{doi:#2}}\else \href
  {http://dx.doi.org/#2} {\textcolor{darkblue}{#1}}\fi \endgroup}
\def\mn@eprint#1#2{\mn@eprint@#1:#2::\@nil}
\def\mn@eprint@arXiv#1{\href {http://arxiv.org/abs/#1} {{\tt arXiv:#1}}}
\def\mn@eprint@dblp#1{\href {http://dblp.uni-trier.de/rec/bibtex/#1.xml}
  {dblp:#1}}
\def\mn@eprint@#1:#2:#3:#4\@nil{\def\@tempa {#1}\def\@tempb {#2}\def\@tempc
  {#3}\ifx \@tempc \@empty \let \@tempc \@tempb \let \@tempb \@tempa \fi \ifx
  \@tempb \@empty \def\@tempb {arXiv}\fi \@ifundefined
  {mn@eprint@\@tempb}{\@tempb:\@tempc}{\expandafter \expandafter \csname
  mn@eprint@\@tempb\endcsname \expandafter{\@tempc}}}

\bibitem[\protect\citeauthoryear{{Abazajian} et~al.,}{{Abazajian}
  et~al.}{2016}]{cmbs4}
{Abazajian} K.~N.,  et~al., 2016, preprint, \href
  {http://adsabs.harvard.edu/abs/2016arXiv161002743A} {} (\mn@eprint {arXiv}
  {1610.02743})

\bibitem[\protect\citeauthoryear{{Beasley}, {Kawata}, {Pearce}, {Forbes}  \&
  {Gibson}}{{Beasley} et~al.}{2003}]{Beasley2003}
{Beasley} M.~A.,  {Kawata} D.,  {Pearce} F.~R.,  {Forbes} D.~A.,   {Gibson}
  B.~K.,  2003, \mndoi [\apjl] {10.1086/379531}, \href
  {http://adsabs.harvard.edu/abs/2003ApJ...596L.187B} {596, L187}

\bibitem[\protect\citeauthoryear{{Brodie} \& {Strader}}{{Brodie} \&
  {Strader}}{2006}]{Brodie2006}
{Brodie} J.~P.,  {Strader} J.,  2006, \mndoi [\araa]
  {10.1146/annurev.astro.44.051905.092441}, \href
  {http://adsabs.harvard.edu/abs/2006ARA%26A..44..193B} {44, 193}

\bibitem[\protect\citeauthoryear{{Dvorkin} \& {Smith}}{{Dvorkin} \&
  {Smith}}{2009}]{dvorkin2009}
{Dvorkin} C.,  {Smith} K.~M.,  2009, \mndoi [\prd]
  {10.1103/PhysRevD.79.043003}, \href
  {http://adsabs.harvard.edu/abs/2009PhRvD..79d3003D} {79, 043003}

\bibitem[\protect\citeauthoryear{{Flesch}}{{Flesch}}{2015}]{Flesch2015}
{Flesch} E.~W.,  2015, \mndoi [\pasa] {10.1017/pasa.2015.10}, \href
  {http://adsabs.harvard.edu/abs/2015PASA...32...10F} {32, e010}

\bibitem[\protect\citeauthoryear{{Forbes}, {Pastorello}, {Romanowsky}, {Usher},
  {Brodie}  \& {Strader}}{{Forbes} et~al.}{2015}]{Forbes2015}
{Forbes} D.~A.,  {Pastorello} N.,  {Romanowsky} A.~J.,  {Usher} C.,  {Brodie}
  J.~P.,   {Strader} J.,  2015, \mndoi [\mnras] {10.1093/mnras/stv1312}, \href
  {http://adsabs.harvard.edu/abs/2015MNRAS.452.1045F} {452, 1045}

\bibitem[\protect\citeauthoryear{{Glover} \& {Clark}}{{Glover} \&
  {Clark}}{2014}]{Glover2014}
{Glover} S.~C.~O.,  {Clark} P.~C.,  2014, \mndoi [\mnras]
  {10.1093/mnras/stt1809}, \href
  {http://adsabs.harvard.edu/abs/2014MNRAS.437....9G} {437, 9}

\bibitem[\protect\citeauthoryear{{Gluscevic}, {Kamionkowski}  \&
  {Hanson}}{{Gluscevic} et~al.}{2013}]{Gluscevic2013}
{Gluscevic} V.,  {Kamionkowski} M.,   {Hanson} D.,  2013, \mndoi [\prd]
  {10.1103/PhysRevD.87.047303}, \href
  {http://adsabs.harvard.edu/abs/2013PhRvD..87d7303G} {87, 047303}

\bibitem[\protect\citeauthoryear{Gorski, Hivon, Banday, Wandelt, Hansen,
  Reinecke  \& Bartelmann}{Gorski et~al.}{2005}]{healpix}
Gorski K.~M.,  Hivon E.,  Banday A.~J.,  Wandelt B.~D.,  Hansen F.~K.,
  Reinecke M.,   Bartelmann M.,  2005, \mndoi [The Astrophysical Journal]
  {10.1086/427976}, 622, 759

\bibitem[\protect\citeauthoryear{{Harris}, {Harris}  \& {Alessi}}{{Harris}
  et~al.}{2013}]{Harris2013}
{Harris} W.~E.,  {Harris} G.~L.~H.,   {Alessi} M.,  2013, \mndoi [\apj]
  {10.1088/0004-637X/772/2/82}, \href
  {http://adsabs.harvard.edu/abs/2013ApJ...772...82H} {772, 82}

\bibitem[\protect\citeauthoryear{{Ishiyama}, {Sudo}, {Yokoi}, {Hasegawa},
  {Tominaga}  \& {Susa}}{{Ishiyama} et~al.}{2016}]{Ishiyama2016}
{Ishiyama} T.,  {Sudo} K.,  {Yokoi} S.,  {Hasegawa} K.,  {Tominaga} N.,
  {Susa} H.,  2016, \mndoi [\apj] {10.3847/0004-637X/826/1/9}, \href
  {http://adsabs.harvard.edu/abs/2016ApJ...826....9I} {826, 9}

\bibitem[\protect\citeauthoryear{{Jacobs} et~al.,}{{Jacobs}
  et~al.}{2016}]{Jacobs2016}
{Jacobs} D.~C.,  et~al., 2016, \mndoi [\apj] {10.3847/0004-637X/825/2/114},
  \href {http://adsabs.harvard.edu/abs/2016ApJ...825..114J} {825, 114}

\bibitem[\protect\citeauthoryear{{Linde}}{{Linde}}{2015}]{linde}
{Linde} A.,  2015, preprint, \href
  {http://adsabs.harvard.edu/abs/2015arXiv151201203L} {} (\mn@eprint {arXiv}
  {1512.01203})

\bibitem[\protect\citeauthoryear{{Mould}}{{Mould}}{1998}]{Mould1998}
{Mould} J.,  1998, \nat, \href
  {http://adsabs.harvard.edu/abs/1998Natur.395A..20M} {395, A20}

\bibitem[\protect\citeauthoryear{{Mutch}, {Geil}, {Poole}, {Angel}, {Duffy},
  {Mesinger}  \& {Wyithe}}{{Mutch} et~al.}{2016}]{dragons}
{Mutch} S.~J.,  {Geil} P.~M.,  {Poole} G.~B.,  {Angel} P.~W.,  {Duffy} A.~R.,
  {Mesinger} A.,   {Wyithe} J.~S.~B.,  2016, \mndoi [\mnras]
  {10.1093/mnras/stw1506}, \href
  {http://adsabs.harvard.edu/abs/2016MNRAS.462..250M} {462, 250}

\bibitem[\protect\citeauthoryear{{Namikawa}}{{Namikawa}}{2017}]{Namikawa2017}
{Namikawa} T.,  2017, preprint, \href
  {http://adsabs.harvard.edu/abs/2017arXiv171100058N} {} (\mn@eprint {arXiv}
  {1711.00058})

\bibitem[\protect\citeauthoryear{{Oesch} et~al.,}{{Oesch}
  et~al.}{2016}]{Oesch2016}
{Oesch} P.~A.,  et~al., 2016, \mndoi [\apj] {10.3847/0004-637X/819/2/129},
  \href {http://adsabs.harvard.edu/abs/2016ApJ...819..129O} {819, 129}

\bibitem[\protect\citeauthoryear{{Omont}, {Willott}, {Beelen}, {Bergeron},
  {Orellana}  \& {Delorme}}{{Omont} et~al.}{2013}]{Omont2013}
{Omont} A.,  {Willott} C.~J.,  {Beelen} A.,  {Bergeron} J.,  {Orellana} G.,
  {Delorme} P.,  2013, \mndoi [\aap] {10.1051/0004-6361/201221006}, \href
  {http://adsabs.harvard.edu/abs/2013A%26A...552A..43O} {552, A43}

\bibitem[\protect\citeauthoryear{{Peng} et~al.,}{{Peng}
  et~al.}{2006}]{Peng2006}
{Peng} E.~W.,  et~al., 2006, \mndoi [\apj] {10.1086/499485}, \href
  {http://adsabs.harvard.edu/abs/2006ApJ...639..838P} {639, 838}

\bibitem[\protect\citeauthoryear{{Pfeffer}}{{Pfeffer}}{2017}]{pfeffer2017}
{Pfeffer} J.,  2017, in prep, 000, 000

\bibitem[\protect\citeauthoryear{{Planck Collaboration} et~al.,}{{Planck
  Collaboration} et~al.}{2016a}]{planck_params}
{Planck Collaboration} et~al., 2016a, \mndoi [\aap]
  {10.1051/0004-6361/201525830}, \href
  {http://adsabs.harvard.edu/abs/2016A%26A...594A..13P} {594, A13}

\bibitem[\protect\citeauthoryear{{Planck Collaboration} et~al.,}{{Planck
  Collaboration} et~al.}{2016b}]{planck_reio}
{Planck Collaboration} et~al., 2016b, \mndoi [\aap]
  {10.1051/0004-6361/201628897}, \href
  {http://adsabs.harvard.edu/abs/2016A%26A...596A.108P} {596, A108}

\bibitem[\protect\citeauthoryear{{Rees}}{{Rees}}{2001}]{rees2001}
{Rees} M.~J.,  2001, ArXiv Astrophysics e-prints, \href
  {http://adsabs.harvard.edu/abs/2001astro.ph..1268R} {}

\bibitem[\protect\citeauthoryear{{Renzini}}{{Renzini}}{2017}]{Renzini2017}
{Renzini} A.,  2017, \mndoi [\mnras] {10.1093/mnrasl/slx057}, \href
  {http://adsabs.harvard.edu/abs/2017MNRAS.469L..63R} {469, L63}

\bibitem[\protect\citeauthoryear{{Roy}, {Lapi}, {Spergel}  \&
  {Baccigalupi}}{{Roy} et~al.}{2018}]{roy18}
{Roy} A.,  {Lapi} A.,  {Spergel} D.,   {Baccigalupi} C.,  2018, preprint, \href
  {http://adsabs.harvard.edu/abs/2018arXiv180102393R} {} (\mn@eprint {arXiv}
  {1801.02393})

\bibitem[\protect\citeauthoryear{{Strader}, {Brodie}, {Cenarro}, {Beasley}  \&
  {Forbes}}{{Strader} et~al.}{2005}]{Strader2005}
{Strader} J.,  {Brodie} J.~P.,  {Cenarro} A.~J.,  {Beasley} M.~A.,   {Forbes}
  D.~A.,  2005, \mndoi [\aj] {10.1086/432717}, \href
  {http://adsabs.harvard.edu/abs/2005AJ....130.1315S} {130, 1315}

\bibitem[\protect\citeauthoryear{{Su}, {Yadav}, {McQuinn}, {Yoo}  \&
  {Zaldarriaga}}{{Su} et~al.}{2011}]{Su2011}
{Su} M.,  {Yadav} A.~P.~S.,  {McQuinn} M.,  {Yoo} J.,   {Zaldarriaga} M.,
  2011, preprint, \href {http://adsabs.harvard.edu/abs/2011arXiv1106.4313S} {}
  (\mn@eprint {arXiv} {1106.4313})

\bibitem[\protect\citeauthoryear{{Turner}}{{Turner}}{2001}]{Turner2001}
{Turner} M.~S.,  2001, ArXiv Astrophysics e-prints, \href
  {http://adsabs.harvard.edu/abs/2001astro.ph..8103T} {}

\bibitem[\protect\citeauthoryear{{Usher} et~al.,}{{Usher}
  et~al.}{2012}]{Usher2012}
{Usher} C.,  et~al., 2012, \mndoi [\mnras] {10.1111/j.1365-2966.2012.21801.x},
  \href {http://adsabs.harvard.edu/abs/2012MNRAS.426.1475U} {426, 1475}

\bibitem[\protect\citeauthoryear{{Vanzella} et~al.,}{{Vanzella}
  et~al.}{2017}]{Vanzella2017}
{Vanzella} E.,  et~al., 2017, \mndoi [\mnras] {10.1093/mnras/stx351}, \href
  {http://adsabs.harvard.edu/abs/2017MNRAS.467.4304V} {467, 4304}

\bibitem[\protect\citeauthoryear{{Venemans} et~al.,}{{Venemans}
  et~al.}{2015}]{Venemans2016}
{Venemans} B.~P.,  et~al., 2015, \mndoi [\mnras] {10.1093/mnras/stv1774}, \href
  {http://adsabs.harvard.edu/abs/2015MNRAS.453.2259V} {453, 2259}

\bibitem[\protect\citeauthoryear{{Wagner-Kaiser} et~al.,}{{Wagner-Kaiser}
  et~al.}{2017}]{Wagner-Kaiser2017}
{Wagner-Kaiser} R.,  et~al., 2017, \mndoi [\mnras] {10.1093/mnras/stx1702},
  \href {http://adsabs.harvard.edu/abs/2017MNRAS.471.3347W} {471, 3347}

\bibitem[\protect\citeauthoryear{{Wang} et~al.,}{{Wang}
  et~al.}{2011}]{Wang2011}
{Wang} R.,  et~al., 2011, \mndoi [\aj] {10.1088/0004-6256/142/4/101}, \href
  {http://adsabs.harvard.edu/abs/2011AJ....142..101W} {142, 101}

\bibitem[\protect\citeauthoryear{{Wang} et~al.,}{{Wang}
  et~al.}{2017}]{Wang2017}
{Wang} L.,  et~al., 2017, preprint, \href
  {http://adsabs.harvard.edu/abs/2017arXiv171007005W} {} (\mn@eprint {arXiv}
  {1710.07005})

\bibitem[\protect\citeauthoryear{{Wiltshire}}{{Wiltshire}}{2009}]{Wiltshire200%
9}
{Wiltshire} D.~L.,  2009, \mndoi [International Journal of Modern Physics D]
  {10.1142/S0218271809016193}, \href
  {http://adsabs.harvard.edu/abs/2009IJMPD..18.2121W} {18, 2121}

\bibitem[\protect\citeauthoryear{{Zaldarriaga} et~al.,}{{Zaldarriaga}
  et~al.}{2008}]{Zaldarriaga2008}
{Zaldarriaga} M.,  et~al., 2008, preprint, \href
  {http://adsabs.harvard.edu/abs/2008arXiv0811.3918Z} {} (\mn@eprint {arXiv}
  {0811.3918})

\bibitem[\protect\citeauthoryear{{da Cunha} et~al.,}{{da Cunha}
  et~al.}{2017}]{taipan}
{da Cunha} E.,  et~al., 2017, \mndoi [\pasa] {10.1017/pasa.2017.41}, \href
  {http://adsabs.harvard.edu/abs/2017PASA...34...47D} {34, e047}

\makeatother
\end{thebibliography}

\begin{acknowledgements}
Parts of this research were conducted by the Australian Research Council Centre of Excellence
for All-Sky Astrophysics under grant CE110001020. FB acknowledges support from an Australian Research Council Future Fellowship (FT150100074). Some of the results in this paper have been derived using the HEALPix \citep{healpix} package.
We thank an anonymous referee for helpful comments.
\end{acknowledgements}

\appendix
\section{Appendix A: Optical depth estimator and power spectrum estimation}
In this appendix, we describe the power spectrum estimation and motivation for the specific form of the estimator in more detail. 

\subsection{Masks}

The first step in calculating a CMB power spectrum is to define the mask on the sky over which the spectrum will be taken. 
We begin by creating a baseline, `full-sky' mask. 
This is the 20\% galactic mask  ($f_{\rm sky}=0.8$) provided by the \textit{Planck} collaboration. 
We apodize the edges with a 2$^\circ$ taper and also mask point sources detected in the high frequency \textit{Planck} channels. 
This mask is used to calculate the full-sky CMB spectrum following the procedure below. 

We also want to create masks that select specific regions of the sky.  
Based on the alleged QSO void morphology,  we construct circular masks of $10^\circ$  radius that  we apodize with a Gaussian taper of FWHM of 30 arcmin. 
These circular masks are multiplied by the full-sky mask in order to avoid the galaxy or bright sources. 
We also impose a restriction that the circles overlap by less than 30\%. 
With this restriction, the size of the circular mask dictates the number of roughly independent patches on the sky. 
We find 72 random locations for $10^\circ$  radius circular patches.

\subsection{Power spectrum estimation}
\label{app:ps-estimation}
For each mask, we measure cross-spectra between the two half-mission foreground-cleaned \texttt{SMICA} CMB temperature maps\footnote{http://irsa.ipac.caltech.edu/data/Planck/release\_2/all-sky-maps/matrix\_cmb.html} (provided in the \texttt{HEALPix}\footnote{http://healpix.sourceforge.net/} format at an angular resolution of 5 arcmin)  to avoid noise debiasing. 
We smoothly bandpass the temperature maps between $200 < \ell < 2000$ on the full-sky before extracting the cross-spectra on individual patches. 
The effect of masking, beam smoothing and pixelation on the recovered spectra is taken into account through the use of the MASTER algorithm (Hivon et al. 2002). 
The reference full-sky power spectrum (the denominator in Eq.~\ref{eq:tau_estimator}) is estimated with the baseline mask above. 
The numerator is calculated from one of the circular masks. 

\subsection{Further details on the estimator} 

There are three points to note about Eq.~\ref{eq:tau_estimator}. 
First, the sum is over a multipole range $\ell \in [300,1900]$. 
The lower end of this range is set by the finite size of the circular patches, while the upper end  is selected based on where instrumental noise and foregrounds become significant. 
Second we have chosen to use $\frac{\ell(\ell+1)}{2\pi} C_\ell$ instead of $C_\ell$ in the sum to avoid over-weighting large angular scales. 
Finally, the estimator uses the ratio of the summed power instead of summing ratios for stability reasons. 

We note that the existence of a positive tail in the real distribution is robust to these analysis choices, although the magnitude of the tail varies. 
The same phenomenology is seen if the radius of the circular regions is reduced to 5$^\circ$; the limited number of patches makes it hard to draw conclusions when the radius is instead increased to 20$^\circ$. 
Positive tails are likewise still seen then the multipole range is varied by either increasing the lower limit to 650 or decreasing the upper limit to 1100 or 1600. 
A positive tail corresponds to lines of sight that undergo earlier reionization. 
\appendix
\section{Appendix B}
We replot Figure 1 in three dimensions in Figure B1, using the distances of \citet{Harris2013} and zooming in on the central 100 Mpc cube.
 \begin{figure*}
\includegraphics[scale=0.7, angle=0]{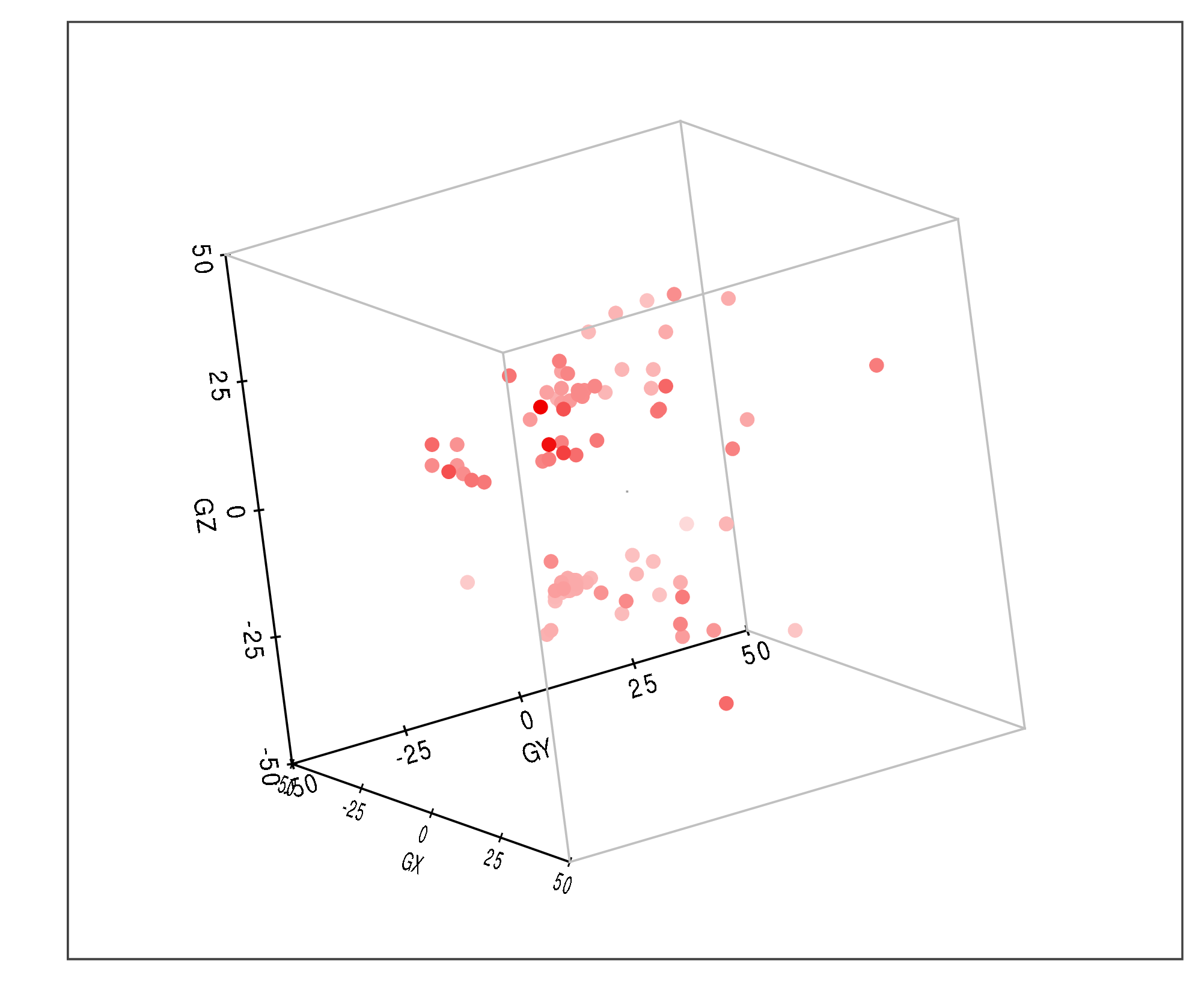}  
\caption{\label{fig:3D}The globular cluster distribution in Galactic cartesian coordinates. GZ = 0 is the Galactic plane. The units are Mpc and the observer is at the origin. Red shading corresponds to depth in the plot. The distribution resembles a filled sphere with radius 40 Mpc after removal of an obscured area in the Galactic plane. This places a lower limit on the size of a local hypothetical population II void.}
\end{figure*}

\end{document}